\documentstyle[aps,floats,prl,twocolumn,psfig]{revtex}
\begin{document} 

\title{ Fermi Surface Evolution, Pseudo Gap and Stagger Gauge\\
Field Fluctuation in Underdoped Cuprates}
\author{Xi Dai and Zhao-Bin Su}
\address{Institute of Theoretical Physics, Chinese Academy of Sciences, Beijing
       100080, P.R.China}
\date{\today}       
\maketitle 

\begin{abstract}
\noindent
 In the context of t-J model we show that in underdoped regime,beside the 
usual long wave length gauge field fluctuation, 
an additional low energy fluctuation, staggered gauge
field fluctuation plays a crucial role in the evolution of Fermi surface(FS)
as well as the line shape of spectral function for the cuprates. 
By including the 
staggered gauge field fluctuation we calculate the spectral function
of the electrons by RPA(random phase approximation).
The line shape of the spectral 
function near $(\pi,0)$
is very broad in underdoped case and is quite sharp in overdoped case.
For the spectral function near $(0.5\pi,0.5\pi)$, the quasiparticle
peaks are always very sharp in both underdoped and overdoped case.
The temperature dependence of the spectral function is also 
discussed in our present calculation. 
These results fit well with the recent ARPES experiments. We also
calculate the FS crossover from a small four segment like FS to a 
large continuous FS. The reason of such kind of FS crossover is ascribed
to the staggered gauge field fluctuation which is strong in underdoped
regime and becomes much weaker in overdoped regime. The pseudo gap
extracted from the ARPES data can be also interpreted by the calculation.

\end{abstract}
\vskip 0.1in
 
%\hspace{1cm}
\vskip 0.1in

The electronic structure of doped cuprates has been
a challenge problem in condensed matter physics\cite{Dag}.  
Recently, angle resolved photoemission spectra(ARPES) experiments have been
done with both underdoped and overdoped samples
\cite{Loe,Mar,Har,HD1,HD2,Nor1,Nor2} . In normal state the 
results show that for
underdoped samples the FS only exists on a small segment near 
$(\pi/2,\pi/2)$\cite{Loe,Mar,Har}.
 Away from this ``Fermi Surface'' segment the line shape
of the spectra is very broad which makes the quasi particle peaks hard
to detect. Also a gap
which is called normal state gap (or pseudo gap) opens there, and
will be closed above a critical temperature 
$T^*$\cite{HD1,Nor1,Nor2}.
But for overdoped samples the FS forms a 
closed curve centered at $(\pi,\pi)$ and the quasi particle peaks are
quite clear in all directions around the FS. The difference in ARPES 
line shape between the 
overdoped and underdoped samples attracts more research interests 
recently\cite{SS,Ch3,Ch4,Nor3}. 

The ARPES results could be explained by several approaches. In the nearly
anti-ferromagnetic fermi liquid(NAFL) theory\cite{NAFL}, the anomalous features 
of the ARPES in underdoped case
can be attributed to short range anti-ferromagnetic fluctuation on a
special kind of FS\cite{SS}. The whole FS is divided into 
two regimes\cite{NAFL} which are called hot particle regime and cold 
particle regime respectively. The hot
particle regime which can be connected by wave vector 
$(\pi,\pi)$ feels the
anti-ferromagnetic fluctuation strongly.It bears the pseudo gap and 
makes the line shape of the spectral function much broader. 
And the cold particle regime can not be connected by the 
wave vector 
$(\pi,\pi)$, so the quasi-particles there only weakly feel 
the anti-ferromagnetic fluctuation and the corresponding 
line shape is much sharper.
Recently
Chubukov {\it et al} have calculated the FS crossover by varying the
 coupling constant from small to large\cite{Ch1}. The small 
FS appears in strong coupling case which can be understood as 
for the underdoped
regime and the large FS is corresponding to weak coupling limit
as for the overdoped regime. We will compare their results
with ours later. 

Another theory which may explain the pseudo gap behavior and FS evolution
is the charge-spin separation scenario. In slave boson approach,
the physical electron is the combination
of the slave boson and pseudo Fermion. The U(1) mean field theory
for the t-J model has been considered by many 
authors\cite{Ub1,MF} using the slave boson
approach. The pseudo gap behavior can be attributed either to d-wave
pairing state or staggered flux state. 
But if the U(1) gauge field fluctuation is considered, the d-wave
pairing state loses its stability\cite{Ub2}.
While for staggered flux phase, 
the missing of translation symmetry and time reverse symmetry has 
been considered as a sort of shortcoming\cite{WL}. In order to 
improve the above defects of the U(1)
mean field theory, P.A.Lee and X.G.Wen proposed the SU(2) mean
field approach for  the t-J model\cite{WL}. In their approach, 
the SU(2)
symmetry is preserved away from half filling by introducing
two kinds of slave bosons. A segment like FS is obtained by
their SU(2) mean field theory.

Inspired by the SU(2) mean field theory and 
our previous study\cite{Dai}, we find
that the staggered flux phase is energetically favorable
 in low doping regime.
So the fluctuation of staggered flux may be important in underdoped regime,
even if there is no symmetry breaking for the 
staggered flux due to the
large quantum fluctuation in 2D .
In order to explore this effect, we introduce a fluctuating 
U(1) staggered gauge field. The time reversal and translational symmetry
get survived because there is no mean field value of staggered
flux. The half pocket like FS is obtained in underdoped regime in which
the staggered gauge fluctuation is strong. With the increment of doping
concentration the staggered gauge fluctuation will be suppressed 
subsequently and meanwhile the FS evolves into a large one. We also
calculate the spectral functions of physical electrons with the staggered
gauge fluctuation being incorporated. The calculated line-shapes
as well as the interpretation for the pseudo-gap are in good 
agreement with the ARPES data.

Follow Lee and Nagaosa\cite{LN}, the effective Lagrangian including the 
phase fluctuation for the uniform RVB state can be written as:

$$
{\cal L}=
\sum_{i\sigma} f_{i\sigma}^+ \left[ \frac{\partial}{\partial\tau}
-\mu_F + i\lambda(r_i) \right] f_{i\sigma}
$$
\begin{equation}
+\sum_{i} b_i^+\left[ 
\frac{\partial}{\partial\tau}-\mu_B+i\lambda(r_i)\right] b_i
\end{equation}
$$
-J'\chi_0\sum_{i,r,\sigma} 
e^{i\theta_{r,i}}f_{i\sigma}^+f_{i+\vec{e}_r,\sigma}
-t\chi_0\sum_{i,r}e^{i\theta_{r,i}} b_i^+b_{i+\vec{e}_r,\sigma}+h.c.
$$
In the above equation, $J'={3 \over 8}J$\cite{Ub1},
the label "r" represents the two directions in 2D plane
and $\vec{e}_r$ represents the two corresponding unit vectors.
Then the three fields
$(\lambda_i,\theta_{1,i},\theta_{2,i})$ can be viewed as the three
components of the U(1) lattice gauge field in 2D. The above 
Lagrangian is invariant under the local U(1) gauge transformation.

In Lee and Nagaosa's approach\cite{LN}, 
only the long wave length part
of the U(1) gauge field is considered. So they treated the above lattice
gauge field Hamiltonian in the long wave length limit and abandoned the 
short wave length part gauge field. Their treatment is understood 
to be valid only in the optimal
doping case. In underdoped case there exists another low energy fluctuation
which is the U(1) gauge field fluctuation near $(\pi,\pi)$. This kind of
low energy fluctuation reflects the instability of staggered flux phase in very
low doping area. With the increasing of doping, the U(1) gauge field
fluctuation near $(\pi,\pi)$ loses its spectral weight rapidly as shown in
Fig.5. Finally at
optimal doping case this kind of fluctuation becomes unimportant and
can be ignore reasonably as in Lee and Nagaosa's paper\cite{LN}. 
In order to include this kind
of fluctuation, we must consider the long wave length expansion of the
gauge field near both $(0,0)$ and $(\pi,\pi)$. In real space, the low
energy modes of the gauge field should include two low energy components,
one is the uniform component and the other is the staggered component.
We have$
\theta_{r,i}={\cal A}_r(\vec{R}_i)+{\cal B}_r(\vec{R}_i)(-1)^{i_x+i_y}
$ with $r=1,2$, and $
\lambda_i={\cal A}_0(\vec{R}_i)+{\cal B}_0(\vec{R}_i)(-1)^{i_x+i_y}
$.

Now a U(1) gauge transformation contains both uniform part
and staggered part:$
\phi_i=\phi_a(\vec{R}_i)+\phi_b(\vec{R}_i)(-1)^{i_x+i_y}
$
The uniform  part of the gauge field is transformed as the usual way:
$
{\cal A}_r(\vec{R})\rightarrow {\cal A}_r(\vec{R})+\partial_r\phi_a(\vec{R})
$ with $r=0,1,2$.
And the stagger part of the gauge field is transformed as:
$
{\cal B}_0(\vec{r}_{mn})\rightarrow {\cal B}_0(\vec{r}_{mn})+\partial_{\tau}\phi_b
$,
$
{\cal B}_r(\vec{R})\rightarrow
{\cal B}_r(\vec{R})-2\phi_b(\vec{R})
$ in which $r=1,2$.

To treat the gauge fluctuation purturbatively, we should expand the 
eq.(1) in $\theta_{ij}$ to the second order. For reason of exploring
the essential physics of the staggered gauge fluctuation, 
in this paper, we 
consider only the effect of staggered gauge field fluctuation. But we will
discuss the effect of uniform gauge field fluctuation whenever 
necessary. 
We choose the gauge fixing condition for the staggered gauge field
as ${\cal B}_1(\vec{r})+{\cal B}_2(\vec{r})=0$.Then in k-space
the Lagrangian with the staggered gauge field fluctuation 
incorporated can be written as
 
$$
H=\sum_{k\sigma}(\epsilon_k-\mu_f) f_{k\sigma}^+f_{k\sigma}+
  \sum_{k}({t \over {J'}}\epsilon_k-\mu_b) b_{k}^+b_{k}
$$
$$
  +\sum_{k,q,\sigma}g(k,q)f_{k+q+Q,\sigma}^+f_{k,\sigma}\varphi_q+
  \sum_{k,q,\sigma}{t \over {J'}}g(k,q)b_{k+q+Q}^+b_{k}\varphi_q
$$
$$
+\sum_{qq'k\sigma}\gamma(k,q,q')f_{k+q+q',\sigma}^+f_{k,\sigma}
 \varphi_q\varphi_{q'}
$$
\begin{equation}
+\sum_{qq'k}{t \over {J'}}\gamma(k,q,q')b_{k+q+q'}^+b_{k}
 \varphi_q\varphi_{q'}
\end{equation}
with $\varphi={\cal B}_1=-{\cal B}_2$,$\epsilon_k=2J'\chi_0
(\cos k_x+\cos k_y)$,
$
g(k,q)=
{J' \over 2}\chi_0\left [e^{-iq_x/2}cos(k_x+q_x/2)-
e^{-iq_y/2}cos(k_y+q_y/2) \right ]
$,$Q=(\pi,\pi)$ and
$
\gamma (k,q,q')=
\frac{\chi_0J'}{16} [e^{-i(q_x+q'_x)/2}\cos (k_x+{q_x \over 2}+{q'_x \over 2}) 
+e^{-i(q_y+q'_y)/2}\cos (k_y+{q_y \over 2}+{q'_y \over 2}) ]
$.

Using eq.(2), we calculate the electronic spectral function in RPA.
First we obtain the effective propagateor of staggered gauge field.
The   stagger gauge field propagator is given by:
$
-{\cal D}^{-1}(q,i\omega_n)=\Pi_f(q,i\omega_n)+\Pi_b(q,i\omega_n)+\Xi_f(q,i\omega_n)+\Xi_b(q,i\omega_n)
$ ,in which $\Pi_f(q,i\omega_n)$ and $\Pi_b(q,i\omega_n)$ are corresponding
to the bubbles of spinons and holons (Fig.1(a) and Fig.1(b)) respectively. And $\Xi_f(q,i\omega_n)$ and
 $\Xi_b(q,i\omega_n)$ are corresponding to Fig.1(c) and Fig.1(d).
The dressed propagator of  spinon and holon are calculated by 
considering the lowest order self energy corrections as shown in Fig.1(e)
and Fig.1(f). Finally we obtain the physical electronic Green's function
by calculating the convolution of spinon and holon's 
Green functions\cite{Dai,LN}. 

\begin{figure}[htb]
%\begin{center}
%\framebox[55mm]{\rule[-21mm]{0mm}{43mm}}
\psfig{file=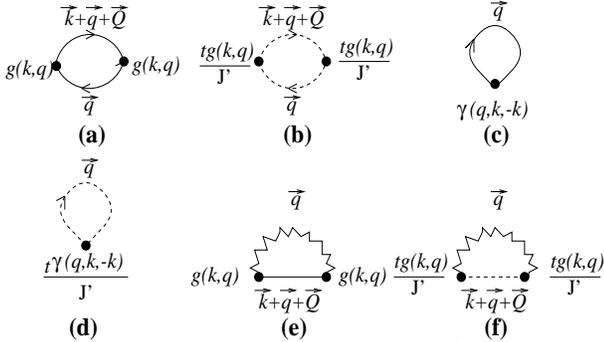,height=45mm,width=80mm,angle=-90}
\caption{~The Feynman diagramm considered in our RPA calculation.
Solid line: spinon Green's function. Dashed line: holon Green's function.
Wagged line: staggered gauge field Green's function. }
\label{fig.1}
%\end{center}
\end{figure}

The spectral weight of the staggered gauge field fluctuation at 
${\vec q}=(0,0)$
of different doping concentration are shown in Fig.2. With the increasing
of doping the staggered gauge field fluctuation loses its spectral weight
rapidly and becomes unimportant when $\delta > 0.15$.

\begin{figure}[htb]
%\begin{center}
%\framebox[55mm]{\rule[-21mm]{0mm}{43mm}}
\psfig{file=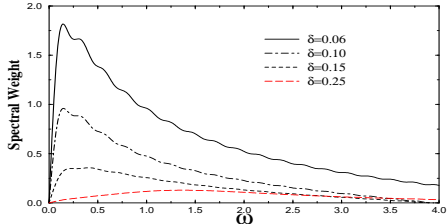,height=35mm,width=70mm,angle=-90}
\caption{~The spectral weight of stagger gauge field at $k=(0,0)$ as a
function of doping concentration. }
%\label{fig:}
%\end{center}
\end{figure}

We have calculated the electronic spectral functions for doping concentration
 $\delta=0.09$ 
and $0.2$ which represent the underdoped and overdoped case respectively.
We choose the parameters as t/J=3 and temperature 
as $T=0.19J$ in our calculation.
For $\delta=0.09$ which is shown in Fig.3(a), 
due to the strong staggered gauge field fluctuation the line shape 
of the spectral
function is very broad and the pseudo gap feature (suppressing the DOS
near the Fermi level) is very clear near $(\pi,0)$. But for the states
near the diagonal line , the line shape is quite sharp and the quasi
particle peaks are well defined here. This is in good agreement with
the recent ARPES studies for underdoped samples\cite{Loe,Mar,Har}.
\begin{figure}[htb]
%\begin{center}
%\framebox[55mm]{\rule[-21mm]{0mm}{43mm}}
\psfig{file=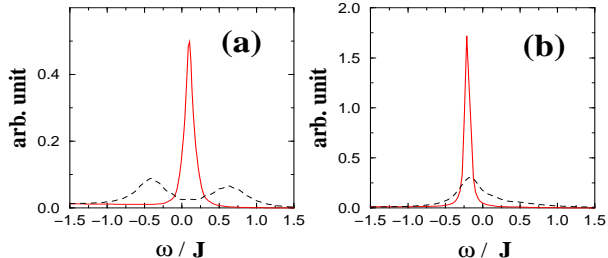,height=35mm,width=80mm,angle=-0}
\caption{~The spectal function of physical electrons for both underdoped(a) and
overdoped case(b). }
%\label{fig}
%\end{center}
\end{figure}
In our present study the large 
difference of the line shape in these two regimes are 
ascribed to the strong k dependent coupling constant $g(k,q)$. 
So unlike the AF fluctuation approach, our results are not
sensitive to the shape of zeroth order FS without 
considering the staggered gauge field fluctuation.
Therefore the essential physics of our interpretation 
should be generic.
The temperature dependence of the spectral function is also obtained
by our RPA calculation. The pseudo-gap depression is only exist 
below a critical temperature $T_{cr}({\vec k})$
which strongly depends on the wave vector ${\vec k}$. The $T_{cr}$
is very high at ${\vec k}=(\pi,0)$, and decrease when
${\vec k}$ is moved toward $(\pi/2,\pi/2)$. The spectral functions
of ${\vec k}=(\pi,0)$ and ${\vec k}=(0.9\pi,0.15\pi)$ 
in two different 
temperature are shown in Fig.4.
\begin{figure}[htb]
%\begin{center}
%\framebox[55mm]{\rule[-21mm]{0mm}{43mm}}
\psfig{file=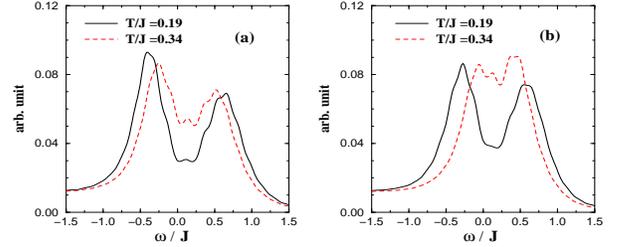,height=35mm,width=80mm,angle=-90}
\caption{The spectral function of electrons for 
 $k=(\pi,0)(a)$ and $k=( 0.9\pi,0.15\pi)(b)$ in two 
different temperature. }
%\label{fig:}
%\end{center}
\end{figure}
 When the temperature is raised 
the pseudo-gap depression is 
absent for ${\vec k}=(0.9\pi,0.15\pi)$(dashed line in Fig.4(b))
 but is still present for
${\vec k}=(\pi,0)$(dashed line in Fig.4(b)). 
These results are also in good agreement with
ARPES experiments\cite{Nor1,Nor2}.

In Fig.3(b)
we give the results in overdoped case. In this case, the staggered 
gauge field fluctuation is unimportant here and it does not change
the spectral function of Lee and Nagaosa's.So the quasi particle peak survives near
$(\pi,0)$, but the width of the peak is much broader than the quasi particle
peaks near $(\pi/2,\pi/2)$. The effect of the uniform gauge field fluctuation
will give an additional width in both the two regime, but the effect of the
uniform gauge field is nearly isotropic and it will not change our result
qualitatively.

Since we can determine the electronic FS by searching for the quasi 
particle peaks moving across the Fermi level\cite{Ch1}, 
a continuous crossover 
from small segment like FS to large
FS is obtained by our calculation. Our results are shown in Fig.5.
In low doping region, the strong staggered gauge
field fluctuation smears the quasi particle peaks dramatically 
near $(\pi,0)$, so we can't find any quasi particle peaks move
across Fermi level in this regime. And in the other limit, along the
direction from $(0,0)$ to $(\pi,\pi)$ the effect of staggered gauge field
fluctuation is quite small,then the quasi particle peaks and the FS
can be detected here. Between the above two limits
there exist a critical point, the FS is present at one 
side and is absent
in the other side.With the increment of doping concentration, this
critical point moves toward $(\pi,0)$. The FS segment becomes
larger and larger and finally the large FS  restores at the optimal
doping which is $15\%$ in our present study. We can compare our approach
with what is used in \cite{Ch1} by Chubukov and {/it et al}. In general the two 
approaches are quite similar. The FS crossover in both approaches
is caused by interacting with a bosonic fluctuation which is strongly
enhanced near the momentum $(\pi,\pi)$.Therefore in our approach, the staggered
gauge field fluctuation plays a similar role with the AF fluctuation
in\cite{Ch1}. But also there exist subtle difference between them.In their approach, 
the coupling constant is nearly isotropic and the pocket like 
FS is obtained only when the next nearest hopping term is included in 
the original Hamiltonian. But in our approach, the coupling constant
$g(k,q)$ is strongly anisotropic. This cause a very large interaction
with the staggered gauge field near $(\pi,0)$ and a quite small interaction
along the diagnose line. So in our study, the unclosed pocket like FS is
caused by the strongly anisotropic coupling constant. 

\begin{figure}[htb]
%\begin{center}
%\framebox[55mm]{\rule[-21mm]{0mm}{43mm}}
\psfig{file=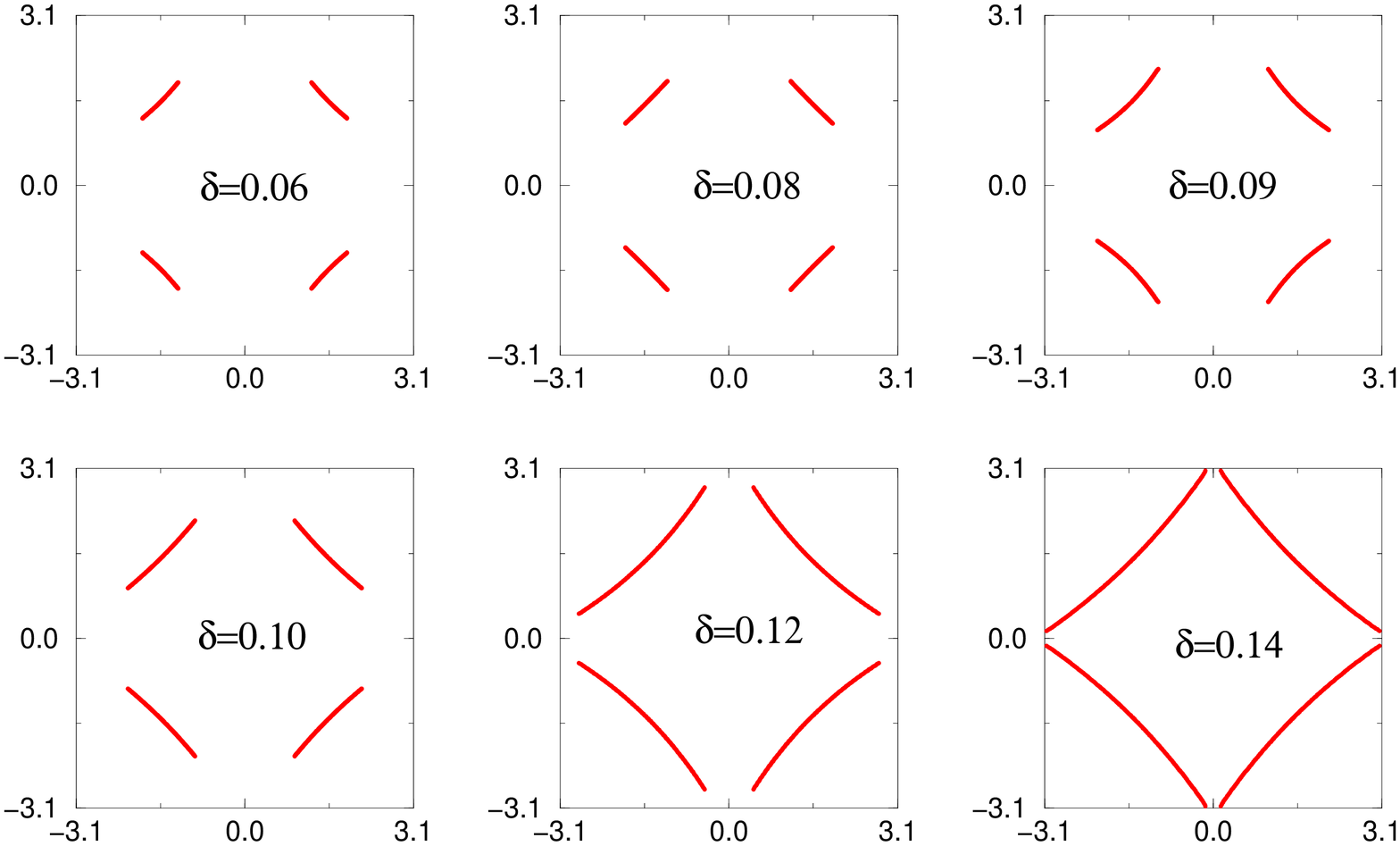,height=55mm,width=80mm,angle=-90}
\caption{~The FS evolution with the increment of doping concentration. }
%\label{fig:}
%\end{center}
\end{figure}

As shown in Fig.3(a) the pseudo gap feature is also very clear 
in the 
electronic spectral function near $(\pi,0)$ in underdoped regime.
We can also calculate the density of states(DOS) of the electrons by
calculating the summation of the spectral function in k space as
$
\rho(\omega)=\sum_kA(k,\omega)
$.
The results are shown in Fig.6(a) for $\delta=6\%$(solid line),$\delta=9\%$
(dashed line) and $\delta=15\%$(dotted line). The suppression of DOS near
Fermi level is clear for $\delta=6\%$ and is absent for $\delta=15\%$. 
This may give a explanation for the pseudo gap behavior in underdoped 
regime. The temperature dependence of the DOS is also calculated and
the results for $\delta=10\%$ are shown in Fig.6(b). One can easily find 
that the pseudo gap is formed with the decrease of the temperature.

In U(1) mean field theory,
the pseudo gap phase is ascribed to either d-wave pairing state or
staggered flux state. The d-wave pairing state only gives four Fermi
point and is known to be unstable against the gauge field 
fluctuation\cite{Ub2}. 
The state which is stable should have `segment' 
like but not `point' like
FS. In the present paper, we proposed a possible description for
the pseudo gap phase in underdoped regime. We
only include the strong fluctuation of staggered
flux and  do not need the symmetry broken
staggered flux which breaks the translation symmetry and time
reverse symmetry. 
We hope this will give a more reasonable explanation for the pseudo
gap behavior in underdoped regime.

\begin{figure}[htb]
%\begin{center}
%\framebox[55mm]{\rule[-21mm]{0mm}{43mm}}
\psfig{file=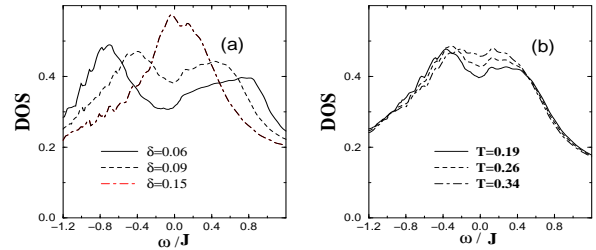,height=35mm,width=80mm,angle=-90}
\caption{~(a)~The density of state(DOS) of physical electrons with different
doping consentration.~(b)~The density of state(DOS) of physical electrons 
with different temperature. }
%\label{fig:}
%\end{center}
\end{figure}

Very recently a phenomenological form of self energy which emphasizes
the d-wave pairing fluctuation\cite{Ioffe} is proposed by M.R.Norman {\it et al}
\cite{Nor2} to 
fit the ARPES experiments.The different behavior in the regime near $(\pi,0)$
and $(\pi/2,\pi/2)$ is ascribed to the symmetry of d-wave pairing. 
As hinted by the SU(2) approach of Lee and Wen, or SO(5) 
theory of Zhang, these two kinds of fluctuation may be 
accommodated together guided by certain underlying subtle physics.

It is rather encouraging that our simple approach fits various kinds
of ARPES data with reasonable magnitude but without 
phenomenological parameters. We expect it catches certain 
essential physics of the pseudo-gap in the under doped regime.

One of the authers(Z. B. Su) would like to thank for the 
helpful discussions
with Prof. Y. K. Bang, A. V. Chubukov, P. A. Lee and L. Yu.

\end{document}